\documentclass[fp,twocolumn]{jpsj3}
\usepackage{txfonts}
\usepackage{color}

\title{Mechanism for Synchronization of Charge Oscillations in Dimer Lattices}

\author{Kenji Yonemitsu$^{1,2}$\thanks{E-mail: kxy@phys.chuo-u.ac.jp} and Philipp Werner$^2$}
\inst{$^1$ Department of Physics, Chuo University, Bunkyo, Tokyo 112-8551, Japan \\
$^2$ Department of Physics, University of Fribourg, 1700 Fribourg, Switzerland} 

\abst{
We discuss the mechanism and the conditions for the appearance of synchronized charge oscillations which have been observed experimentally and theoretically after strong photoexcitation of dimerized systems. In the Hubbard model with on-site repulsion, the Bloch equations for a wave-number-dependent pseudospin -- whose components describe the charge-density difference, current density, and bond density between the two sublattices -- involve an alternatingly tilted pseudomagnetic field, which assists the synchronization of pseudospins with different wave numbers, irrespective of the initial condition. This fact is numerically confirmed by the dynamics in finite lattices based on the exact diagonalization method. In the presence of nearest-neighbor repulsion, however, the synchronization can be hindered by excitons. Therefore, the excitation of a sufficiently large density of free electron-hole pairs, but low density of excitons, is needed to achieve synchronization. 
}

\begin{document}
\maketitle

\section{Introduction}
Among various nonequilibrium phenomena, the many-electron dynamics in far-from-equilibrium lattice systems has attracted much attention.\cite{basov_rmp11,nicoletti_aop16,giannetti_aip16,kaiser_ps17,chergui_cr17,kawakami_jpb18,ishihara_jpsj19} Many interesting effects, including photoinduced phase transitions, are nonlinear and caused by interactions.\cite{iwai_prl03,fausti_s11,wall_nphys11,tsuji_prl11,tsuji_prb12,gao_n13,ohara_prb13,ishikawa_ncomms14,kaiser_prb14,stojchevska_s14,fukaya_ncomms15,han_sciadv15,mitrano_n16,rettig_ncomms16,singer_prl16,mor_prl17,ono_prl17,murakami_prl17,tanaka_prb18,kinoshita_prl20,gillmeister_ncomms20,schueler_prx20,bittner_prb20,werner_prb20} With progress in the experimental techniques, ultrafast phenomena that necessarily involve high-energy processes become increasingly  important. Which electronic phase is realized in a correlated electron material is often determined by a subtle competition between the kinetic and interaction terms in the Hamiltonian. Thus, low-energy properties are linked to high-energy processes, which makes ultrafast control possible in some cases. 

During photoexcitation, high harmonic generation is observed in solids and reflects properties of their electronic states.\cite{ghimire_nphys11,schubert_np14,luu_n15,yoshikawa_s17,silva_np18,murakami_prl18,ikeda_pra18,nag_prb19,lysne_prb20b} Recently, second harmonic generation has been observed in a centrosymmetric organic superconductor $\kappa$-(bis[ethylenedithio]tetrathiafulvalene)$_2$Cu[N(CN)$_2$]Br [$\kappa$-(BEDT-TTF)$_2$Cu[N(CN)$_2$]Br] through a nonlinear petahertz current before substantial scattering processes occur.\cite{kawakami_ncomms20} In this compound,\cite{kagawa_n05,kawakami_prl09,abdel_prb10,naka_jpsj10,gomi_prb10,hotta_prb10,dayal_prb11,yonemitsu_jpsj11b,itoh_prl13,gomi_jpsj14} which is known to have a dimer lattice, stimulated emission is observed after strong photoexcitation.\cite{kawakami_np18} This stimulated emission is also caused by a nonlinear charge oscillation -- an electronic breathing mode -- as has been shown by numerical calculations based on the exact diagonalization method.\cite{yonemitsu_jpsj18a} When the photoexcitation is weak, i.e., when the optical field amplitude is small, different charge oscillations appear whose frequencies correspond to the peak energies in the optical conductivity spectrum. When photoexcitation is strong, however, these charge oscillations synchronize to produce the electronic breathing mode. The synchronization can be demonstrated numerically by introducing randomness into the transfer integrals to suppress it and showing that sufficiently strong on-site repulsion overcomes the randomness and recovers the effect.\cite{shimada_jpsj20} It has also been discussed that strong on-site attraction produces a pair analog of the electronic breathing mode.\cite{yonemitsu_jpsj18b} 

A widely studied lattice with two atoms in a unit cell is the honeycomb lattice, on which a synchronization transition and resultant coherent oscillations have been reported to occur in a recent mean-field investigation.\cite{nag_prb19} Here, it was pointed out that the equations of motion are somewhat similar to those in the Kuramoto model,\cite{kuramoto_book84,acebron_rmp05,rodrigues_pr16} where the presence of a synchronization transition is established. This analogy is highly suggestive, but it implies that whether or not the in-phase synchronization is realized depends on the sign of the interaction. Hence, the detailed form of the equations of motion merits further investigations. In the present work, we will adopt a pseudospin representation and show that the synchronization is indeed sensitive to the sign of the interaction, a conclusion that will be confirmed numerically using the exact diagonalization method. We hope that these insights into the mechanism and conditions for the synchronization of charge oscillations will be helpful for future experiments related to ultrafast control of charge motion. 

Within a mean-field approximation, charge-oscillation dynamics is described by the Bloch equations for pseudospins in a similar manner to those for photoinduced magnetization\cite{tsuji_prl13} and pairing\cite{barankov_prl04} dynamics, or photoinduced excitonic condensation dynamics,\cite{murakami_prl17} which allows us to intuitively grasp the essence of the collective evolution. The time-dependent BCS pairing problem is known to be integrable,\cite{barankov_prl04} and its similarity to and difference from the present problem needs to be clarified. On short timescales, the mean-field picture basically holds even when electron correlations are taken into account.\cite{tsuji_prl13,werner_prb20} 

The numerical calculations in this paper consider one-dimensional dimerized lattices. They are bipartite and consist of two sublattices. However, the synchronization mechanism described by the Bloch equations is independent of the dimensionality of the system, and the synchronization phenomena have been numerically observed in one- and two-dimensional systems.\cite{yonemitsu_jpsj18a,shimada_jpsj20,yonemitsu_jpsj18b} As representative systems, we treat the Hubbard and a spinless fermion model. The former represents a model with repulsive interactions within a sublattice, while the latter possesses repulsive interactions between the sublattices, and their synchronization conditions are different. To show the sensitivity to the initial condition in the latter model, we prepare initial nonequilibrium states in different ways, by photoexcitation and quenching. 

\section{Pseudospin Representation}
To describe collective dynamics, the pseudospin representation is often useful. Here, following Ref.~\citen{nag_prb19}, we employ a mean-field approximation to treat charge-oscillation dynamics in this representation and investigate the mechanism for synchronization. We consider a Hubbard model with on-site repulsion $U$ and a spinless fermion model with nearest-neighbor repulsion $V$ on dimer lattices. Although the discussion in this section does not depend on the dimensionality of the system, we use one-dimensional lattices with alternating transfer integrals, $t_1$ and $t_2$, and periodic boundary conditions when the kinetic term is explicitly shown, to make the relationship with the results in the next section clear. The distance between neighboring sites is set to be equal and unity when the system is photoexcited.\cite{yonemitsu_jpsj18a} The Hubbard model can be written as 
\begin{eqnarray}
H_U & = & \sum_{j=0}^{N-1} \sum_\sigma \left[ 
t_1 \left( e_{j\sigma}^\dagger o_{j\sigma}     
+          o_{j\sigma}^\dagger e_{j\sigma} \right) \right.\nonumber\\
&&\quad+\left. 
t_2 \left( o_{j\sigma}^\dagger e_{j+1\sigma} 
+        e_{j+1\sigma}^\dagger o_{j\sigma} \right) 
\right.\nonumber \\
& &\quad + \left. U \left(
e_{j\uparrow}^\dagger e_{j\uparrow} 
e_{j\downarrow}^\dagger e_{j\downarrow} + 
o_{j\uparrow}^\dagger o_{j\uparrow} 
o_{j\downarrow}^\dagger o_{j\downarrow}
\right) \right]
\;, \label{eq:Hubbard}
\end{eqnarray}
while the Hamiltonian of the spinless fermion model is  
\begin{eqnarray}
H_V & = & \sum_{j=0}^{N-1} \left[ 
t_1 \left( e_j^\dagger o_j     +     o_j^\dagger e_j \right) + 
t_2 \left( o_j^\dagger e_{j+1} + e_{j+1}^\dagger o_j \right) 
\right. \nonumber \\
& & \quad + \left. V \left(
e_j^\dagger e_j o_j^\dagger o_j + o_j^\dagger o_j e_{j+1}^\dagger e_{j+1}
\right) \right]
\;, \label{eq:spinless}
\end{eqnarray}
where $ e_{j\sigma}^\dagger $ ($ e_{j}^\dagger $) creates an electron with spin $\sigma$ (a fermion) at site $j$ of the ``even'' sublattice and  $ o_{j\sigma}^\dagger $ ($ o_{j}^\dagger $) creates an electron with spin $\sigma$ (a fermion) at site $j$ of the ``odd'' sublattice. 

In what follows in this section, we mainly discuss the Hubbard model and its charge-oscillation dynamics. When nontrivial differences between the two models exist, they will be pointed out. We use a mean-field approximation corresponding to the Hartree approximation, which allows us to capture the essence of interaction effects. Via the Fourier transforms 
$ e_{k\sigma} = (1/\sqrt{N}) \sum_j e^{-ikj} e_{j\sigma} $ and 
$ o_{k\sigma} = (1/\sqrt{N}) \sum_j e^{-ikj} o_{j\sigma} $ 
with $N$ denoting the number of dimers, the mean-field Hamiltonian can be written as 
\begin{equation}
H_U^{\mbox{MF}} = \sum_{k\sigma}
\begin{pmatrix} e_{k\sigma}^\dagger & o_{k\sigma}^\dagger \\ \end{pmatrix}
\begin{pmatrix}
(U/2)\delta n & h(k) \\
h^\ast(k)     & -(U/2)\delta n \\
\end{pmatrix}
\begin{pmatrix}
e_{k\sigma} \\ o_{k_\sigma} \\
\end{pmatrix}
\;, \label{eq:mf_Hubbard}
\end{equation}
with 
\begin{equation}
\delta n = \langle e_{j\sigma}^\dagger e_{j\sigma} 
- o_{j\sigma}^\dagger o_{j\sigma} \rangle 
\;, \label{eq:deln_Hubbard}
\end{equation}
for the Hubbard model and as 
\begin{equation}
H_V^{\mbox{MF}} = \sum_k
\begin{pmatrix} e_k^\dagger & o_k^\dagger \\ \end{pmatrix}
\begin{pmatrix}
-2V \delta n & h(k) \\
h^\ast(k)    & 2V \delta n \\
\end{pmatrix}
\begin{pmatrix}
e_k \\ o_k \\
\end{pmatrix}
\;, \label{eq:mf_spinless}
\end{equation}
with 
\begin{equation}
\delta n = \frac12 \langle e_j^\dagger e_j 
-  o_j^\dagger o_j \rangle 
\;, \label{eq:deln_spinless}
\end{equation}
for the spinless fermion model, where constant terms have been omitted. On the one-dimensional lattice, the off-diagonal element is given by 
\begin{equation}
h(k) = t_1 + t_2 e^{-ik} 
\;. \label{eq:dispersion}
\end{equation}

Following Ref.~\citen{nag_prb19}, we define 
\begin{equation}
\begin{pmatrix} e_{k\sigma}^\dagger & o_{k\sigma}^\dagger \\ \end{pmatrix} = 
\begin{pmatrix} a_{k\sigma}^\dagger & b_{k\sigma}^\dagger \\ \end{pmatrix} 
\frac{1}{\sqrt{2}}
\begin{pmatrix} 1 & -1 \\ 1 & 1 \\ \end{pmatrix}
\begin{pmatrix} e^{-i\phi_k/2} & 0 \\ 0 & e^{i\phi_k/2} \\ \end{pmatrix}
\;, \label{eq:unitary}
\end{equation}
with $ \phi_k=\tan^{-1} \left[ \mbox{Im}\left( h(k) \right)
/\mbox{Re} \left( h(k) \right) \right] $ and
rewrite the mean-field Hamiltonian in Eq.~(\ref{eq:mf_Hubbard}) as 
\begin{equation}
H_U^{\mbox{MF}} = \sum_{k\sigma}
\begin{pmatrix} a_{k\sigma}^\dagger & b_{k\sigma}^\dagger \\ \end{pmatrix}
\begin{pmatrix}
-h^\prime(k) & (U/2)\delta n \\
(U/2)\delta n   & h^\prime(k) \\
\end{pmatrix}
\begin{pmatrix}
a_{k\sigma} \\ b_{k_\sigma} \\
\end{pmatrix}
\;, \label{eq:mf_Hubbard_ab}
\end{equation}
with $ h^\prime(k) = \mbox{sgn} \, \mbox{Re} \left( h(0) \right) | h(k) | $ and $ \delta n $ in Eq.~(\ref{eq:deln_Hubbard}) as 
\begin{equation}
\delta n = \frac{1}{2N} \sum_{k\sigma} \langle a_{k\sigma}^\dagger b_{k\sigma} + b_{k\sigma}^\dagger a_{k\sigma} \rangle 
\;. \label{eq:deln_Hubbard_ab}
\end{equation}
As in Ref.~\citen{barankov_prl04}, we define the pseudospin components as 
\begin{equation}
r_{1k\sigma} = \langle a_{k\sigma}^\dagger b_{k\sigma} + 
b_{k\sigma}^\dagger a_{k\sigma} \rangle
\;, \label{eq:ps_1}
\end{equation}
\begin{equation}
r_{2k\sigma} = \langle -i a_{k\sigma}^\dagger b_{k\sigma} +i
b_{k\sigma}^\dagger a_{k\sigma} \rangle
\;, \label{eq:ps_2}
\end{equation}
\begin{equation}
r_{3k\sigma} = \langle a_{k\sigma}^\dagger a_{k\sigma} -
b_{k\sigma}^\dagger b_{k\sigma} \rangle
\;. \label{eq:ps_3}
\end{equation}
Here the normalization condition 
$ r_{1k\sigma}^2 + r_{2k\sigma}^2 + r_{3k\sigma}^2 =1 $ 
is always satisfied for $k\sigma$ for which one state is initially occupied and the other is initially unoccupied, and $ r_{1k\sigma}^2 + r_{2k\sigma}^2 + r_{3k\sigma}^2 =0 $ is satisfied otherwise. In equilibrium with $ \delta n = 0 $ and $ h^\prime(k) < 0 $ (i.e., $ t_1+t_2 < 0 $), $ r_{3k\sigma}=-1 $ for $k\sigma$ in the former case. Further, we define $\Omega$\cite{barankov_prl04} as $ \Omega = (U/2)\delta n $. The self-consistency condition leads to 
\begin{equation}
\Omega = \frac{U}{4N} \sum_{k\sigma} r_{1k\sigma}
\;. \label{eq:sceq}
\end{equation}
The pseudospin dynamics is described by the Bloch equations,
\begin{equation}
\dot{r}_{1k\sigma} = 2h^\prime(k) r_{2k\sigma} 
\;, \label{eq:bloch_1}
\end{equation}
\begin{equation}
\dot{r}_{2k\sigma} = -2h^\prime(k) r_{1k\sigma} -2\Omega r_{3k\sigma} 
\;, \label{eq:bloch_2}
\end{equation}
\begin{equation}
\dot{r}_{3k\sigma} = 2\Omega r_{2k\sigma}
\;, \label{eq:bloch_3}
\end{equation}
which are consistent with the normalization condition. For $ U > 0 $, there is no static solution with $ \Omega \neq 0 $. 

For the spinless fermion model with $ \Omega \equiv 2V\delta n = (V/N) \sum_k r_{1k} $, the Bloch equations can be obtained by replacing $ \Omega $ by $ -\Omega $ in Eqs.~(\ref{eq:bloch_1})-(\ref{eq:bloch_3}). For $ V $ larger than a critical value $ V_\text{c} $, there is a static solution with $ \Omega \neq 0 $, and the dynamical problem becomes equivalent to the time-dependent BCS pairing problem.\cite{barankov_prl04} 

\begin{figure}
\includegraphics[width=\columnwidth]{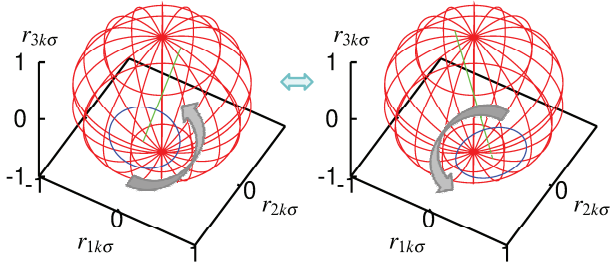}
\caption{(Color online)
Precession of a pseudospin when the majority of the pseudospins have positive $x$ components (left panel) and negative $x$ components (right panel) for $ U > 0 $. 
\label{fig:two_spheres_arrows}}
\end{figure}

Equations~(\ref{eq:bloch_1})-(\ref{eq:bloch_3}) describe the Larmor precession of the pseudospin 
\begin{equation}
\dot{ \mbox{\boldmath $r$} }_{k\sigma}(t) = \mbox{\boldmath $B$}_k(t) 
\times \mbox{\boldmath $r$}_{k\sigma}(t)
\;, \label{eq:bloch_v}
\end{equation}
in a pseudo magnetic field 
\begin{equation}
\mbox{\boldmath $B$}_k(t) = ^{t}\!\left( B_{1k}(t), B_{2k}, B_{3k} \right) = 
^{t}\!\left( 2\Omega(t), 0, -2h^\prime(k) \right)
\;, \label{eq:mag_field}
\end{equation}
where the time dependence is made explicit here, and $ B_{3k} = -2h^\prime(k) > 0 $ for $ h^\prime(k) < 0 $. The $x$, $y$, and $z$ components of the pseudospin correspond to the charge-density difference, current density, and bond density between the ``even'' and ``odd'' sublattices. The rate of rotation around the $z$ axis depends on the wave number. In what follows, we assume that $ U > 0 $. The rotation axis is tilted by $ \Omega(t) $ whose sign is that of $ \delta n(t) $, i.e., that of the majority of $ r_{1k\sigma}(t) $. When the majority of $ r_{1k\sigma}(t) $ take positive (negative) values, $ \Omega(t) $ is positive (negative), and the rotation axis is tilted to have a positive (negative) $x$ component, as shown in the left (right) panel of Fig.~\ref{fig:two_spheres_arrows}. 
By tilting the rotation axis in this way, most of the pseudospins are located on the same side of the rotation axis, those near the bottom $^{t}(0, 0, -1)$ are accelerated, and consequently their dynamics become similar, which assists the synchronization of charge oscillations. These considerations suggest that charge oscillations are synchronized regardless of the initial distribution of $ r_{1k\sigma} $. 

On the right hand side of Eq.~(\ref{eq:bloch_2}), the sign of $ h^\prime(k) $ and that of $ r_{3k\sigma} $ are the same, as is obvious from Eqs.~(\ref{eq:mf_Hubbard_ab}) and (\ref{eq:ps_3}), at least near equilibrium and the sign of the majority of $ r_{1k\sigma}(t) $ and the sign of $ \Omega(t) $ are the same, as already pointed out. Thus, a $U$-driven (see Eq.~\eqref{eq:sceq}) but $k$-independent force component is applied in the direction of the majority of the $k$-dependent current flow, which is equivalent to the above picture based on the tilted rotation axis. 

A solution to the Bloch equations can be obtained, according to Ref.~\citen{barankov_prl04}, by the ansatz 
\begin{equation}
r_{1k\sigma}=A_k \Omega\;,\;\;\;\;
r_{2k\sigma}=B_k \dot{\Omega}\;,\;\;\;\;
r_{3k\sigma}=C_k \Omega^2 -D_k
\;. \label{eq:ansatz}
\end{equation}
The Bloch equations and the normalization condition lead to
$ A_k=2h^\prime(k) B_k $, $ B_k=C_k $, and 
\begin{equation}
\dot{\Omega}^2 +\Omega^4 
+\left( (2h^\prime(k))^2-\frac{2D_k}{C_k} \right)\Omega^2 
+ \frac{D_k^2-1}{C_k^2}=0
\;. \label{eq:k_dep_dif_eq}
\end{equation}
If the above equation is independent of $k$, and of the form
\begin{equation}
\dot{\Omega}^2 +\Omega^4 
+\left( \Delta_-^2 -\Delta_+^2 \right)\Omega^2 
-\Delta_-^2 \Delta_+^2 =0
\;, \label{eq:k_indep_dif_eq}
\end{equation}
it has a solution described by the elliptic cosine function, 
\begin{equation}
\Omega(t)=\Delta_+ \mbox{cn} \left(
\sqrt{ \Delta_+^2+\Delta_-^2 }t, 
\frac{\Delta_+}{\sqrt{ \Delta_+^2+\Delta_-^2 }}
\right)
\;. \label{eq:ellliptic_cn}
\end{equation}
When we set the amplitude $ \Delta_+=\delta $ and the frequency 
$ \sqrt{ \Delta_+^2+\Delta_-^2 }=-2h^\prime(0) $ for $ t_1, t_2 < 0 $\cite{yonemitsu_jpsj18a}, the elliptic modulus becomes 
$\Delta_+/\sqrt{ \Delta_+^2+\Delta_-^2 }=\delta/(-2h^\prime(0)) $. In order for Eq.~(\ref{eq:k_dep_dif_eq}) to become Eq.~(\ref{eq:k_indep_dif_eq}), the $k$-dependent factors must satisfy 
\begin{equation}
C_k=-2
\left[ \left( \left( 2h^\prime(0) \right)^2-\left( 2h^\prime(k) \right)^2 \right)^{2}+4\left( 2h^\prime(k) \right)^{2}\delta^2 \right]^{-\frac12}
\;, \label{eq:c_k}
\end{equation}
\begin{equation}
D_k=\left[ \left( 2h^\prime(0) \right)^{2} 
-\left( 2h^\prime(k) \right)^{2} -2\delta^2 \right] 
\frac{C_k}{-2}
\;. \label{eq:d_k}
\end{equation}
The negative sign in Eq.~(\ref{eq:c_k}) is due to the self-consistency condition, Eq.~(\ref{eq:sceq}), which can be rewritten as $ 1/U=1/(4N)\sum_{k\sigma}^\prime 2h^\prime(k) C_k $, where the summation is over the $k\sigma$ for which the pseudospin has a nonzero magnitude. In the limit of vanishing oscillation amplitude $ \delta \rightarrow 0 $, $ D_k \rightarrow 1 $, and $ r_{3k\sigma} \rightarrow -1 $, i.e., the solution indeed approaches the equilibrium state. Although the initial condition must satisfy Eqs.~(\ref{eq:c_k}) and (\ref{eq:d_k}) for the pseudospin dynamics to be exactly described by the elliptic cosine function and this is not generally achieved by a photoexcitation, this solution strongly suggests that charge oscillations are synchronized by a sufficiently large $U$, 
which has indeed been observed previously\cite{yonemitsu_jpsj18a,shimada_jpsj20} and will also be demonstrated in the next section. 
The electronic breathing mode is more easily described in real space rather than in momentum space, but it is controlled by a set of equations analogous to Eqs.~(\ref{eq:bloch_1})-(\ref{eq:bloch_3}), as is explained in the Appendix.

For the spinless fermion model, the situation is quite different from the Hubbard model. For $ V > V_\text{c} $, a static charge ordered state is realized in equilibrium. As mentioned before, its nonequilibrium dynamics is known\cite{barankov_prl04} to be described by another elliptic function, the delta amplitude, whose small-amplitude limit corresponds to the Higgs amplitude mode. For $ 0 < V < V_\text{c} $, the excitonic effect stems from the coupling among electron-hole pairs $a_{k}^\dagger b_{k}$ with different $k$'s through the element $2V\delta n$ of the mean-field Hamiltonian. An electron and a hole are bound by $V$ to form an exciton, which is described by a linear combination of $a_{k}^\dagger b_{k}$. If the initial state contains such excitons, the charge-oscillation dynamics contains a slow component and the synchronization is difficult to achieve. In the pseudospin representation, the rotation axis is tilted in the direction opposite to the Hubbard case. Thus, a $V$-driven $k$-independent force component is applied in the direction opposite to the current flow, which allows the existence of a charge-ordered state for $ V > V_\text{c} $. For $ 0 < V < V_\text{c} $, however, this $k$-independent force component may lead to a synchronization if the initial state does not have a significant excitonic component. This fact is numerically demonstrated in the next section. 

Note that a similar synchronization phenomenon of pseudospins has been discussed in the context of the Higgs mode in superconductors.\cite{matsunaga_s14,tsuji_prb15} There, the Bloch equations were linearized with respect to the time-dependent parts, from which a resonant precession of pseudospins was obtained in the long-time asymptotic evolution of the order parameter. In the present case, on the other hand, the equilibrium state has no order parameter. After the strong photoexcitation, the system has no external field and its initial state is far from equilibrium. Although Eq.~(\ref{eq:c_k}) has a factor that appears associated with resonance, the synchronization mechanism is different from that in the resonant precession discussed in Ref.~\citen{tsuji_prb15}. 

\section{Dynamics from Charge Disproportionation}
Employing one-dimensional lattices, Eqs.~(\ref{eq:Hubbard}) and (\ref{eq:spinless}), with periodic boundary conditions, we numerically solve the time-dependent Schr\"odinger equation to investigate the charge-oscillation dynamics from far-from-equilibrium states using the exact diagonalization method.\cite{yonemitsu_prb09} For the transfer integrals, we use $t_1$=$-$0.3 and $t_2$=$-$0.1, from which the frequency of the electronic breathing mode is given by\cite{yonemitsu_jpsj18a}
\begin{equation}
\omega_\text{osc} = 2(| t_1 | + | t_2 |)=0.8 
\;.\label{eq:omega_osc}
\end{equation}
Note that this frequency is independent of $U$ in the Hubbard model and independent of $V$ in the one-dimensional spinless fermion and extended Hubbard models, although it is lowered by intersite repulsion $V_{ij}$ in the two-dimensional extended Hubbard model for $\kappa$-(BEDT-TTF)$_2$Cu[N(CN)$_2$]Br.\cite{yonemitsu_jpsj18a} In the noninteracting case, the electronic breathing mode is merely one of the optically active modes in dimer lattices. The other modes are suppressed by $U$ or $V$ after strong photoexcitation. 
The time slice employed in the numerical solutions of the time-dependent Schr\"odinger equation is $ dt $=0.02 for small $U$ or $V$, while smaller $ dt $ values are used for large $U$ or $V$ to ensure the conservation of the total energy and of the norm. The number of sites is denoted by $L$ ($L=2N$). We use system sizes up to $L$=16 for the Hubbard model at quarter filling and up to $L$=24 for the spinless fermion model at half filling. When photoexcitations are used to produce nonequilibrium states, the procedure is the same as in previous studies\cite{yonemitsu_jpsj18a,shimada_jpsj20}, i.e., we use symmetric one-cycle electric-field pulses, $ A(t) = (cF/\omega_c) \left[ \cos (\omega_c t)-1 \right] \theta(t+(2\pi/\omega_c)) \theta(-t) $, with central frequency $ \omega_c $ and field amplitude $F$. In the quench calculations, we add a staggered potential $ -\Delta \sum_{j\sigma}\left( e_{j\sigma}^\dagger e_{j\sigma} -o_{j\sigma}^\dagger o_{j\sigma}  \right) $ to Eq.~(\ref{eq:Hubbard}) or $ -\Delta \sum_{j}\left( e_{j}^\dagger e_{j} -o_{j}^\dagger o_{j}  \right) $ to Eq.~(\ref{eq:spinless}) to obtain the ground state and use this state as the initial state at $t$=0. The charge-oscillation dynamics is then computed for $t>0$ with $\Delta$=0. Fourier spectra of the charge density denote the absolute values of the Fourier transforms of the time profiles ($0 < t < 10^3$) of the charge density immediately after setting $ \Delta $=0 or immediately after the photoexcitation. 

Before showing numerical results, we consider the Hubbard model in the limit of infinitely large repulsion, $U=\infty$. In this limit, it becomes equivalent to a spinless fermion model at doubled filling without interaction, 
\begin{equation}
H_{V=0} = \sum_k
\begin{pmatrix} a_k^\dagger & b_k^\dagger \\ \end{pmatrix}
\begin{pmatrix}
-h^\prime(k) & 0           \\
0            & h^\prime(k) \\
\end{pmatrix}
\begin{pmatrix}
a_k \\ b_k \\
\end{pmatrix}
\;, \label{eq:spinless_V=0}
\end{equation}
with $ h^\prime(k)=-\sqrt{t_1^2+t_2^2+2t_1t_2\cos k} $, i.e., to an integrable system; thus, independent of the frequency, the charge oscillations do not decay. At quarter filling, $(N/2)$ spin-up and $(N/2)$ spin-down electrons are present. When an electron is virtually moved through a distance of $L$ sites, it passes over $(N/2)$ fermions with the opposite spin. If $(N/2)$ is odd, the outcome is different from the similar process in the noninteracting spinless fermion model or in the fully polarized Hubbard model by a factor $(-1)$. Thus, the allowed wave numbers are $k=(2\pi/N)j$ with $j$ being an integer for even $(N/2)$ and $k=(\pi/N)(2j+1)$ for odd $(N/2)$. On the other hand, the Hubbard model with $U$=0 is equivalent to the noninteracting spinless fermion model at the same filling; thus, the allowed wave numbers are always $k=(2\pi/N)j$. The charge-density difference $2\delta n$ can be rewritten as 
\begin{equation}
2\delta n 
= \frac{1}{N}\sum_j \langle e_j^\dagger e_j -o_j^\dagger o_j \rangle 
= \frac{1}{N}\sum_k \langle a_k^\dagger b_k +b_k^\dagger a_k \rangle
\;. \label{eq:deln_spinless_ab}
\end{equation}
Thus, it is maximized by the state $ \prod_k \frac{1}{\sqrt{2}}\left( a_k^\dagger +b_k^\dagger \right) | 0 \rangle $, which is set to be the initial state of $ | \Psi(t) \rangle $, so that
\begin{equation}
| \Psi(t) \rangle = \prod_k \frac{1}{\sqrt{2}}\left( e^{i h^\prime(k) t} a_k^\dagger +e^{-i h^\prime(k) t} b_k^\dagger \right) | 0 \rangle
\;. \label{eq:Psi}
\end{equation}
In this state, the time evolution of $2\delta n$ is given by 
\begin{equation}
2\delta n(t) = \frac{1}{N} \sum_k \cos 2h^\prime(k) t
\;. \label{eq:deln_spinless_evol}
\end{equation}
Its Fourier spectrum has peaks at $ \omega_k = 2\sqrt{t_1^2+t_2^2+2t_1t_2\cos k} $ with the $k$ values described above, i.e., $k$=0, $\pi$ for $U$=0 and $\pm \pi/2$ for $U$=$\infty$ if $L$=4 ($N$=2), $k$=0, $\pm \pi/2$, $\pi$ for $U$=0 and $U$=$\infty$ if $L$=8 ($N$=4), etc. 

\subsection{Hubbard model at quarter filling}
\begin{figure}
\includegraphics[width=\columnwidth]{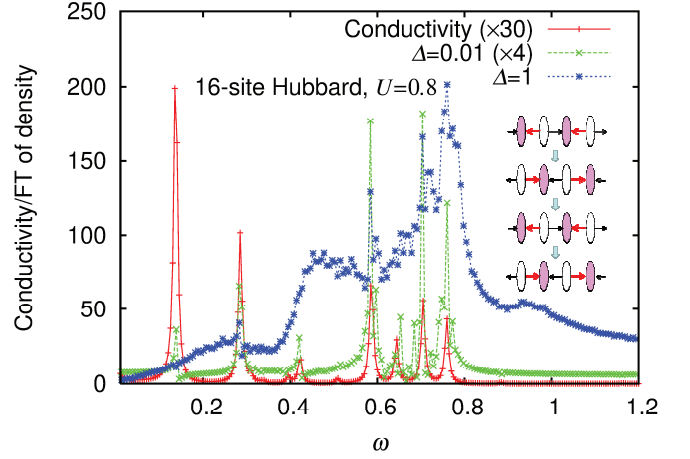}
\caption{(Color online)
Optical conductivity spectrum in the ground state and Fourier spectra of the charge density from near-equilibrium ($\Delta$=10$^{-2}$) and far-from-equilibrium ($\Delta$=1) initial states in the Hubbard model with $L$=16 and $U$=0.8. 
The inset illustrates the breathing mode on the dimerized lattice. 
\label{fig:ky_1d16_FRsigma_wndu0p8}}
\end{figure}

First we show in Fig.~\ref{fig:ky_1d16_FRsigma_wndu0p8} that the Fourier spectrum of the charge density, $ \sum_\sigma \langle e_{j\sigma}^\dagger e_{j\sigma} \rangle $ or $ \sum_\sigma \langle o_{j\sigma}^\dagger o_{j\sigma} \rangle $, evolving from a near-equilibrium initial state, has peaks at energies where the optical conductivity spectrum has peaks. 
Here, the optical conductivity spectrum is calculated for the ground state as in previous studies.\cite{yonemitsu_jpsj11b} It is also shown in this figure that the Fourier spectrum of the charge density evolving from the near-equilibrium state with small charge disproportionation ($\Delta$=10$^{-2}$) is quite different from that evolving from a state with large charge disproportionation ($\Delta$=1). The latter is dominated by the electronic breathing mode at $\omega$=0.8, although this mode is significantly broadened, reflecting its short lifetime. 

\begin{figure}
\includegraphics[width=\columnwidth]{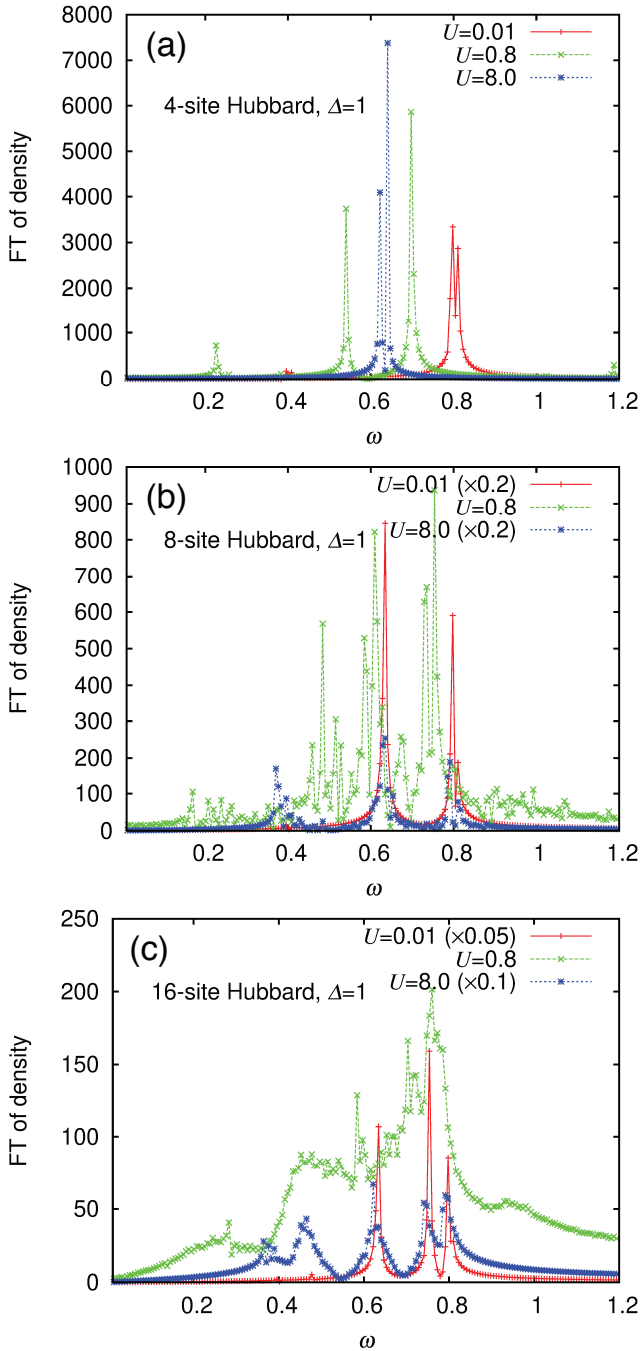}
\caption{(Color online)
Fourier spectra of the charge density from far-from-equilibrium ($\Delta$=1) initial states in the Hubbard model with (a) $L$=4, (b) $L$=8, (c) $L$=16 for $U$=10$^{-2}$, 0.8, and 8.0. 
\label{fig:ky_1dLx_FRwnd1000e1ux}}
\end{figure}

The Fourier spectra of the charge density evolving from the state with large charge disproportionation ($\Delta$=1) are shown in Figs.~\ref{fig:ky_1dLx_FRwnd1000e1ux}(a) to \ref{fig:ky_1dLx_FRwnd1000e1ux}(c) for different system sizes from $L$=4 to $L$=16. 
For very weak repulsion ($U$=10$^{-2}$), the spectra have discrete peaks whose energies are given by $\omega_k$ with $k=(2\pi/N)j$. For very strong repulsion ($U$=8.0), the systems become close to the noninteracting spinless fermion model with periodic (for even $N/2$) or antiperiodic (for odd $N/2$) boundary conditions; thus, the spectra have discrete peaks whose energies are given by $\omega_k$ with $k=(2\pi/N)j$ or $k=(\pi/N)(2j+1)$, respectively. In both cases, the systems are close to integrable ones, and all the charge oscillations are long-lived, i.e., the peaks are narrow and high. In the figures for larger systems, the corresponding peak heights are multiplied by factors smaller than unity, which allows them to be compared with intermediate repulsion ($U$=0.8) cases. Correlation effects become significant when the repulsion strength is intermediate. In this case, charge oscillations are synchronized\cite{yonemitsu_jpsj18a,shimada_jpsj20} until they finally decay due to dephasing. 
The resultant electronic breathing mode has a broad peak near $\omega$=$ \omega_\text{osc} $. The peak position is slightly below $ \omega_\text{osc} $ for $L$=16, but it rapidly approaches $ \omega_\text{osc} $ with increasing $L$, and it is expected to be at $ \omega_\text{osc} $ in the thermodynamic limit. 
As the system size $L$ increases, the difference between the very weak/strong and the intermediate repulsion cases, i.e., the difference between the nearly integrable and the strongly correlated cases, becomes apparent in Fig.~\ref{fig:ky_1dLx_FRwnd1000e1ux}. 

\begin{figure}
\includegraphics[width=\columnwidth]{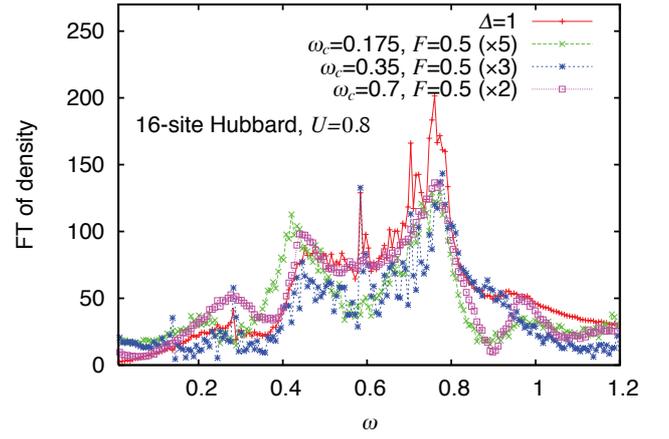}
\caption{(Color online)
Fourier spectra of the charge density from far-from-equilibrium initial states prepared by $\Delta$=1 and by strong ($F$=0.5) photoexcitations with $\omega_c$=0.175, 0.35, and 0.7 in the Hubbard model with $L$=16 and $U$=0.8. 
\label{fig:ky_1d16_FRq_vs_phu0p8}}
\end{figure}

For the largest system ($L$=16) with intermediate repulsion ($U$=0.8) we have calculated the Fourier spectra with different initial states, as shown in 
Fig.~\ref{fig:ky_1d16_FRq_vs_phu0p8}. 
The initial states with large charge disproportionation are obtained either as the ground state with large alternating site energies ($\Delta$=1) or by the application of one-cycle large-amplitude ($F$=0.5) electric-field pulses with different central frequencies $\omega_c$. Although the detailed structures are different, all the spectra are similar and dominated by the electronic breathing mode at $\omega$=0.8. Thus, the synchronization of charge oscillations is independent of the preparation of the initial state. It is universally observed for states with large charge disproportionation in the Hubbard model with intermediate repulsion.

\subsection{Spinless fermion model at half filling}
\begin{figure}
\includegraphics[width=\columnwidth]{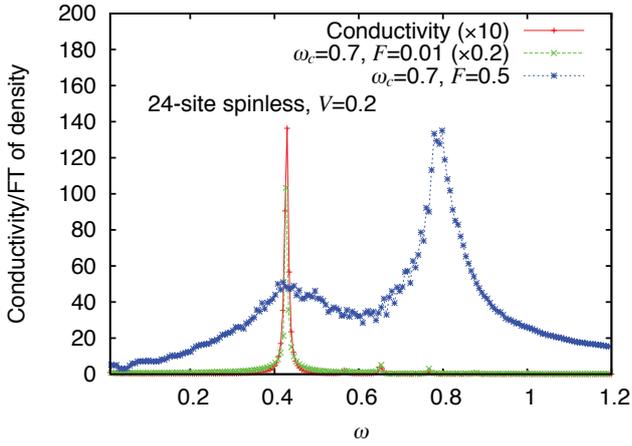}
\caption{(Color online)
Optical conductivity spectrum in the ground state and Fourier spectra of the charge density after weak ($F$=10$^{-2}$) and strong ($F$=0.5) photoexcitations with $\omega_c$=0.7 in the spinless fermion model with $L$=24 and $V$=0.2. 
\label{fig:ky_1dsf24_FRsigma_bwndv0p2}}
\end{figure}

In the previous section, we mentioned that the excitonic effect should be taken into account in the model with nearest-neighbor repulsion. This implies that the charge-oscillation dynamics is sensitive to how the initial state is prepared. Hence, we mainly use one-cycle electric-field pulses with a central frequency near the absorption edge at $\omega$=0.8, $\omega_c$=0.7, which allow excitations of many free electron-hole pairs. We show in Fig.~\ref{fig:ky_1dsf24_FRsigma_bwndv0p2} that the Fourier spectrum of the charge density, $ \langle e_{j}^\dagger e_{j} \rangle $ or $ \langle o_{j}^\dagger o_{j} \rangle $, evolving from a near-equilibrium state, has peaks at energies where the optical conductivity spectrum has peaks. 
Both spectra are dominated by an excitonic peak at $\omega \simeq$0.4 even if the near-equilibrium initial state is prepared by a weak photoexcitation ($F$=10$^{-2}$) with $\omega_c$=0.7. This figure also shows that, if the initial state is prepared by a strong photoexcitation ($F$=0.5) to be far from equilibrium, the Fourier spectrum is dominated by the electronic breathing mode at $\omega$=0.8, which is in contrast to the Fourier spectrum after a weak photoexcitation. 

\begin{figure}
\includegraphics[width=\columnwidth]{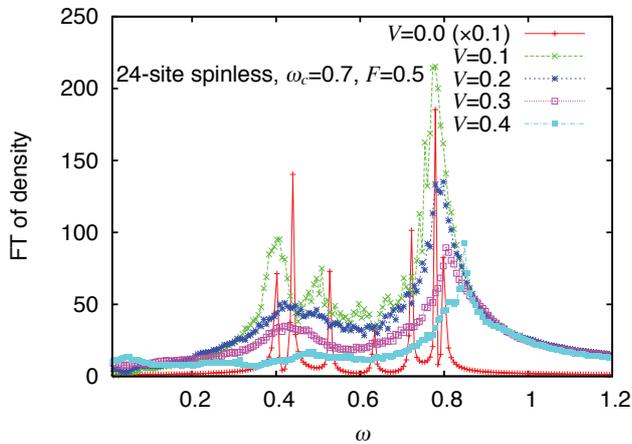}
\caption{(Color online)
Fourier spectra of the charge density after strong ($F$=0.5) photoexcitations with $\omega_c$=0.7 in the spinless fermion model with $L$=24 for $V$ values below $ V_\text{c} $. 
\label{fig:ky_1dsf24_FRbwnd1009f0p5_vx}}
\end{figure}

The Fourier spectra of the charge density after strong photoexcitations ($F$=0.5) are shown in Fig.~\ref{fig:ky_1dsf24_FRbwnd1009f0p5_vx} for different repulsion strengths $V$ below the critical value $ V_\text{c} $. 
For $V$=0, the system is integrable, and all the charge oscillations have infinitely long lifetimes. The corresponding discrete peaks appear at $\omega_k$ with $k$=$(\pi/6)j$, and their peak heights are multiplied by 0.1 in order for them to be comparable with the nonintegrable cases. For small $V$ ($V$=0.1), some structures in the Fourier spectrum reflect the peaks at $V$=0, but they are significantly broadened. For $V$=0.1, 0.2, and 0.3, the Fourier spectra are dominated by the electronic breathing mode at $\omega$=0.8, but its peak height decreases with increasing $V$. For $V$=0.4, the corresponding peak appears slightly blueshifted, although $V$ is still smaller than $ V_\text{c} $. In general, the peak associated with the electronic breathing mode is considerably broadened when $V$ approaches $ V_\text{c} \simeq 2| t_1 |$=0.6, and it disappears in the charge-ordered phase at $ V > V_\text{c} $.\cite{yonemitsu_jpsj18a} 

\begin{figure}
\includegraphics[width=\columnwidth]{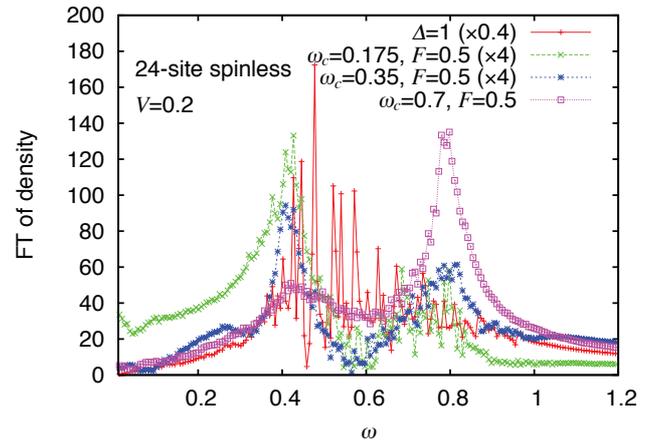}
\caption{(Color online)
Fourier spectra of the charge density from far-from-equilibrium initial states prepared by $\Delta$=1 and by strong ($F$=0.5) photoexcitations with $\omega_c$=0.175, 0.35, and 0.7 in the spinless fermion model with $L$=24 and $V$=0.2. 
\label{fig:ky_1dsf24_FRq_vs_phv0p2}}
\end{figure}

For $V$=0.2, we have also calculated the Fourier spectra with different initial states with large charge disproportionation, as shown in Fig.~\ref{fig:ky_1dsf24_FRq_vs_phv0p2}. 
They are sensitive to the initial condition. When the initial state is prepared by a photoexcitation with $\omega_c$=0.7, the spectrum is dominated by the electronic breathing mode. However, when $\omega_c$ is lowered to allow the exciton at $\omega \simeq$0.4 (Fig.~\ref{fig:ky_1dsf24_FRsigma_bwndv0p2}) to be excited, the Fourier spectra have large contributions from the exciton and smaller contributions from the electronic breathing oscillation. When the initial sate is prepared by $\Delta$=1, the Fourier spectrum becomes more complex, presumably because various electron-hole pairs are involved. Thus, to synchronize charge oscillations, the initial state must be strongly photoexcited with a frequency that produces many free and few bound electron-hole pairs. 

\section{Conclusions}
To clarify the mechanism for the previously observed synchronization of charge oscillations\cite{kawakami_np18,yonemitsu_jpsj18a,shimada_jpsj20} and investigate the conditions for the appearance of this phenomenon, we analytically studied the Bloch equations in the mean-field approximation and numerically studied the charge-oscillation dynamics using the exact diagonalization method for the Hubbard and spinless fermion models on one-dimensional dimerized lattices. 

In one of the Bloch equations, the time derivatives of current densities between the sublattices are determined by two terms, one of which is of kinetic origin and the other is of interaction origin. The kinetic term depends on the wave number, while the term derived from the interaction is proportional to a wave number independent and thus universal factor. 
The latter facilitates the synchronization of charge oscillations. In the Hubbard model with $U>0$, more precisely speaking, when the interaction within a sublattice is repulsive, the kinetic term and the interaction term have the same sign and constructively work together. From the viewpoint of pseudospins, the rotation axis describing the Larmor precession is alternatingly tilted in such a way that it leads pseudospins of different wave numbers to be synchronized. In the spinless fermion model with $V>0$, more precisely speaking, when the interaction between the two sublattices is repulsive, the kinetic term and the interaction term have opposite signs. This allows the existence of static charge order for $ V > V_\text{c} $. For $ V < V_\text{c} $, the synchronization of charge oscillations is possible but requires suitable excitation protocols.

The above insights based on the analytic form of the Bloch equations are all consistent with the Fourier analyses of the numerically obtained charge-oscillation dynamics. To prepare initial states with charge disproportionation, we either quenched a staggered potential to zero or applied one-cycle electric-field pulses. In the Hubbard model with intermediate repulsion, i.e., sufficiently away from the integrable limits of $U$=0 and $U$=$\infty$, the synchronization is achieved irrespective of how the initial state is prepared. In the spinless fermion model with $ V < V_\text{c} $, a sufficiently large number of free electron-hole pairs must be excited to achieve the synchronization, which is hindered by excitons and is not achieved by the quenching. These results clarify the conditions for the synchronization of charge oscillations on dimer lattices, which may guide the study of synchronization phenomena in different classes of lattices. 

\begin{acknowledgments}
This work was supported by JSPS KAKENHI Grant No. JP16K05459,                          MEXT Q-LEAP Grant No. JPMXS0118067426, and JST CREST Grant No. JPMJCR1901. PW acknowledges support from ERC Consolidator Grant No.~724103.
\end{acknowledgments}

\appendix
\section{Breathing Mode in the Pseudospin Picture}
\begin{figure}
\centering
\includegraphics[width=0.7\columnwidth]{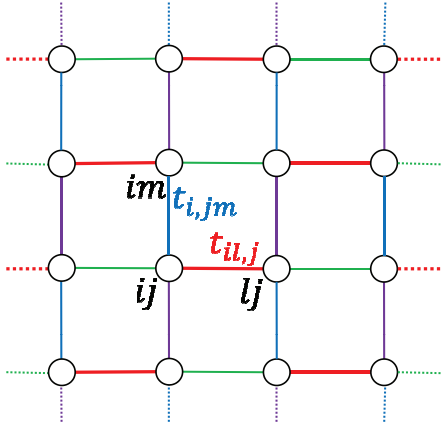}
\caption{(Color online)
Notation used for the transfer integrals between sites on the square lattice. 
\label{fig:square_lattice}}
\end{figure}

In this appendix we show that the equations governing the charge motion in a real-space representation are equivalent to the pseudo-spin equations (\ref{eq:bloch_1})-(\ref{eq:bloch_3}) and explain the synchronization leading to the observed breathing mode. 
Although the following discussion does not depend on the dimensionality of the system, we consider the electronic breathing mode in the square lattice shown in Fig.~\ref{fig:square_lattice}.
The Hubbard model can be written as 
\begin{eqnarray}
H_\text{2D} & = & 
\sum_{ilj\sigma} t_{il,j} c_{ij\sigma}^\dagger c_{lj\sigma} + 
\sum_{ijm\sigma} t_{i,jm} c_{ij\sigma}^\dagger c_{im\sigma} \nonumber\\
&& + U \sum_{ij} 
c_{ij\uparrow}^\dagger   c_{ij\uparrow} 
c_{ij\downarrow}^\dagger c_{ij\downarrow} 
\;, \label{eq:sq_lattice}
\end{eqnarray}
where $ c_{ij\sigma}^\dagger $ creates an electron with spin $\sigma$ at site ($i$,$j$). In the Heisenberg picture, the time evolution of the one-body operator $ c_{ij\sigma}^\dagger c_{lm\sigma} $ is given by 
\begin{eqnarray}
\frac{d}{dt} c_{ij\sigma}^\dagger c_{lm\sigma} & = & 
-i\sum_{l'} t_{ll',m} c_{ij\sigma}^\dagger c_{l'm\sigma} 
-i\sum_{m'} t_{l,mm'} c_{ij\sigma}^\dagger c_{lm'\sigma} \nonumber\\
&& 
+i\sum_{i'} t_{ii',j} c_{i'j\sigma}^\dagger c_{lm\sigma}
+i\sum_{j'} t_{i,jj'} c_{ij'\sigma}^\dagger c_{lm\sigma} \nonumber\\
&& -iU\left( 
c_{lm\bar{\sigma}}^\dagger c_{lm\bar{\sigma}} 
-c_{ij\bar{\sigma}}^\dagger c_{ij\bar{\sigma}}
\right) c_{ij\sigma}^\dagger c_{lm\sigma}
\;. \label{eq:sq_eq_motion}
\end{eqnarray}
Using Eq.~(\ref{eq:sq_eq_motion}), we easily obtain 
\begin{eqnarray}
&& \frac{d}{dt} \left( 
c_{ij\sigma}^\dagger c_{lm\sigma}
-c_{ab\sigma}^\dagger c_{cd\sigma}
\right) \nonumber\\
&& = -i\left( 
\sum_{l'} t_{ll',m} c_{ij\sigma}^\dagger c_{l'm\sigma}+
\sum_{m'} t_{l,mm'} c_{ij\sigma}^\dagger c_{lm'\sigma} 
\right) \nonumber\\
&& \quad -i\left( 
\sum_{a'} t_{aa',b} c_{a'b\sigma}^\dagger c_{cd\sigma}+
\sum_{b'} t_{a,bb'} c_{ab'\sigma}^\dagger c_{cd\sigma} 
\right) \nonumber\\
&& \quad +i\left( 
\sum_{i'} t_{ii',j} c_{i'j\sigma}^\dagger c_{lm\sigma}+
\sum_{j'} t_{i,jj'} c_{ij'\sigma}^\dagger c_{lm\sigma} 
\right) \nonumber\\
&& \quad +i\left( 
\sum_{c'} t_{cc',d} c_{ab\sigma}^\dagger c_{c'd\sigma}+
\sum_{d'} t_{c,dd'} c_{ab\sigma}^\dagger c_{cd'\sigma} 
\right) \nonumber\\
&& \quad -iU\left( 
c_{lm\bar{\sigma}}^\dagger c_{lm\bar{\sigma}} 
-c_{ij\bar{\sigma}}^\dagger c_{ij\bar{\sigma}}
\right) c_{ij\sigma}^\dagger c_{lm\sigma} \nonumber\\
&& \quad +iU\left( 
c_{cd\bar{\sigma}}^\dagger c_{cd\bar{\sigma}} 
-c_{ab\bar{\sigma}}^\dagger c_{ab\bar{\sigma}}
\right) c_{ab\sigma}^\dagger c_{cd\sigma} 
\;. \label{eq:bloch_r1}
\end{eqnarray}
When we assume that the sites ($i$,$j$) and ($l$,$m$) belong to the even sublattice and the sites ($a$,$b$) and ($c$,$d$) belong to the odd sublattice, the sites ($a'$,$b$), ($a$,$b'$), ($c'$,$d$), and ($c$,$d'$) belong to the even sublattice, and the sites ($i'$,$j$),($i$,$j'$),($l'$,$m$),and ($l$,$m'$) belong to the odd sublattice. Thus, the first four terms can be regarded as current densities between the two sublattices weighted by transfer integrals. In the last two terms, we use the mean-field approximation to replace the one-body operators with spin $\bar{\sigma}$=$-\sigma$ by their expectation values. Because the sites ($i$,$j$) and ($l$,$m$) [($a$,$b$) and ($c$,$d$)] belong to the same sublattice, the last two terms vanish owing to the translational symmetry. Thus, Eq.~(\ref{eq:bloch_r1}) is equivalent to Eq.~(\ref{eq:bloch_1}) for the charge-density difference between the sublattices. Using Eq.~(\ref{eq:sq_eq_motion}), we also obtain 
\begin{eqnarray}
&& \frac{d}{dt} \left( 
-ic_{ij\sigma}^\dagger c_{lm\sigma}
+ic_{lm\sigma}^\dagger c_{ij\sigma}
\right) \nonumber\\
&& = -\sum_{l'} t_{ll',m} \left( 
c_{ij\sigma}^\dagger c_{l'm\sigma} +
c_{l'm\sigma}^\dagger c_{ij\sigma} 
\right) \nonumber\\
&& \quad -\sum_{m'} t_{l,mm'} \left( 
c_{ij\sigma}^\dagger c_{lm'\sigma} +
c_{lm'\sigma}^\dagger c_{ij\sigma} 
\right) \nonumber\\
&& \quad +\sum_{i'} t_{ii',j} \left( 
c_{i'j\sigma}^\dagger c_{lm\sigma} +
c_{lm\sigma}^\dagger c_{i'j\sigma} 
\right) \nonumber\\
&& \quad +\sum_{j'} t_{i,jj'} \left( 
c_{ij'\sigma}^\dagger c_{lm\sigma} +
c_{lm\sigma}^\dagger c_{ij'\sigma} 
\right) \nonumber\\
&& \quad + U\left( 
c_{ij\bar{\sigma}}^\dagger c_{ij\bar{\sigma}}-
c_{lm\bar{\sigma}}^\dagger c_{lm\bar{\sigma}}
\right)\left(
c_{ij\sigma}^\dagger c_{lm\sigma} +
c_{lm\sigma}^\dagger c_{ij\sigma}
\right)
\;, \label{eq:bloch_r2}
\end{eqnarray}
and
\begin{eqnarray}
&& \frac{d}{dt} \left( 
c_{ij\sigma}^\dagger c_{lm\sigma}+
c_{lm\sigma}^\dagger c_{ij\sigma}
\right) \nonumber\\
&& = \sum_{l'} t_{ll',m} \left( 
-ic_{ij\sigma}^\dagger c_{l'm\sigma} 
+ic_{l'm\sigma}^\dagger c_{ij\sigma} 
\right) \nonumber\\
&& +\sum_{m'} t_{l,mm'} \left( 
-ic_{ij\sigma}^\dagger c_{lm'\sigma} 
+ic_{lm'\sigma}^\dagger c_{ij\sigma} 
\right) \nonumber\\
&& -\sum_{i'} t_{ii',j} \left( 
-ic_{i'j\sigma}^\dagger c_{lm\sigma} 
+ic_{lm\sigma}^\dagger c_{i'j\sigma} 
\right) \nonumber\\
&& -\sum_{j'} t_{i,jj'} \left( 
-ic_{ij'\sigma}^\dagger c_{lm\sigma} 
+ic_{lm\sigma}^\dagger c_{ij'\sigma} 
\right) \nonumber\\
&& -U\left( 
c_{ij\bar{\sigma}}^\dagger c_{ij\bar{\sigma}}-
c_{lm\bar{\sigma}}^\dagger c_{lm\bar{\sigma}}
\right)\left(
-ic_{ij\sigma}^\dagger c_{lm\sigma}
+ic_{lm\sigma}^\dagger c_{ij\sigma}
\right)
\;. \label{eq:bloch_r3}
\end{eqnarray}
When we assume that the site ($i$,$j$) belongs to the even sublattice and the site ($l$,$m$) belongs to the odd sublattice, the sites ($l'$,$m$) and ($l$,$m'$) belong to the even sublattice, and the sites ($i'$,$j$) and ($i$,$j'$) belong to the odd sublattice. Then, the first and second terms in Eqs.~(\ref{eq:bloch_r2}) and (\ref{eq:bloch_r3}) are one-body operators acting within the even sublattice, and the third and fourth terms are ones acting within the odd sublattice. The first four terms in Eq.~(\ref{eq:bloch_r2}) can be regarded as charge-density differences between the two sublattices weighted by transfer integrals. 
In the mean-field approximation, $ U\left( c_{ij\bar{\sigma}}^\dagger c_{ij\bar{\sigma}}-c_{lm\bar{\sigma}}^\dagger c_{lm\bar{\sigma}} \right) $ is replaced by $ U\langle c_{ij\bar{\sigma}}^\dagger c_{ij\bar{\sigma}}-c_{lm\bar{\sigma}}^\dagger c_{lm\bar{\sigma}} \rangle = U\delta n = 2\Omega $, which is independent of ($i$,$j$), ($l$,$m$), or $\bar{\sigma}$. The last term in Eq.~(\ref{eq:bloch_r2}) is regarded as ($-2\Omega$) times the bond density between sites ($i$,$j$) and ($l$,$m$). Thus, Eq.~(\ref{eq:bloch_r2}) is equivalent to Eq.~(\ref{eq:bloch_2}) for the current density between the sublattices. The first and second (third and fourth) terms in Eq.~(\ref{eq:bloch_r3}) can be regarded as current densities within the even (odd) sublattice weighted by transfer integrals, which give a small contribution and are unimportant. [In momentum space, they can be eliminated by the unitary transformation within each sublattice using $ e^{\mp i\phi_k/2} $ in Eq.~(\ref{eq:unitary}).] The last term in Eq.~(\ref{eq:bloch_r3}) can be regarded as ($-2\Omega$) times the current density between sites ($i$,$j$) and ($l$,$m$). Thus, Eq.~(\ref{eq:bloch_r3}) is equivalent to Eq.~(\ref{eq:bloch_3}) [though both sides are multiplied by ($-1$)] for the bond density between the sublattices. 

It is now clear that a common factor $U\delta n$(=$2\Omega$) appears independently of the relative position of sites ($i$,$j$) and ($l$,$m$). When $U$ is positive, the force in the last term of Eq.~(\ref{eq:bloch_r2}) is applied in the direction of the current flow. [The current direction is easily checked by setting ($l$,$m$)=($i$,$j$) in Eq.~(\ref{eq:sq_eq_motion}).] Thus, the charge-density difference between the sublattices synchronizes the charge motion on different bonds to achieve the electronic breathing mode. 

\bibliography{s_mech}

\end{document}